\documentclass[aps,prb,twocolumn,showpacs,preprintnumbers,amsmath,amssymb,superscriptaddress,floatfix]{revtex4}
\usepackage{graphics,bm}
\usepackage{amssymb}
\usepackage{epsfig}
\usepackage{epsf}
\begin{document}

\title{Theory of reduced superfluid density in underdoped cuprate superconductors}
\author{Wei-Cheng Lee}
\affiliation{Department of Physics, The University of Texas at Austin, Austin, TX 78712, USA}
\email{leewc@mail.utexas.edu}
\author{Jairo Sinova}
\affiliation{Department of Physics, Texas A\&M University, College Station, TX 77843-4242, USA}
\author{A. A. Burkov}
\affiliation{Department of Physics and Astronomy, University of Waterloo, Waterloo, Ontario, Canada N2L 3G1}
\author{Yogesh Joglekar}
\affiliation{Department of Physics, Indiana University-Purdue University Indianapolis, Indianapolis, Indiana 46202, USA}
\author{A. H. MacDonald}
\affiliation{Department of Physics, The University of Texas at Austin, Austin, TX 78712, USA}

\date{\today}

\begin{abstract}
The critical temperature of an underdoped cuprate superconductor is limited 
by its phase stiffness $\rho$. 
In this article we argue that the dependence of $\rho$ 
on doping $x$ should be understood as a consequence of deleterious 
competition with antiferromagnetism at large electron densities, rather than as 
evidence for pairing of holes in the $x =0$ Mott insulator state.
$\rho$ is suppressed at small $x$ because the correlation energy of a $d$-wave superconductor
has a significant pairing-wavevector dependence when antiferromagnetic fluctuations are strong. 
\end{abstract}

\maketitle
%\vskip1pc

%\newpage

\section{Introduction} 

The fascinating and rich phenomenology of 
high temperature cuprate superconductors has been very thoroughly studied 
over the past 20 years.  Although there is substantial variability in detail
from material to material, all cuprates exhibit robust Mott insulator antiferromagnetism 
when the hole-doping fraction $x$ is very small, superconductivity which
appears when $x$ exceeds a minimum value $\sim 0.1$, and a maximum 
$T_c$ in optimally doped materials with $x \sim 0.2$.  In the underdoped 
regime, the superconducting transition temperature 
is limited by phase fluctuations\cite{uemura,kivelson,ubc,ohiostate}, and experiments 
hint at a wide variety of (typically) short-range correlations
associated with competing charge and spin orders.
The underdoped regime poses a fundamental challenge to theory because
its electronic properties are not fully consistent with any of the various well-understood 
{\em fixed-point} behaviors that often help us to classify and predict the
properties of very complex materials.  

The phenomenological parameter $\rho$ used to characterize phase-fluctuation stiffness 
in a superconductor is normally expressed in terms of the superfluid density $n_s$ by 
writing $\rho = \hbar^2 n_s/m^*$, an identification that is partly justified by 
BCS mean-field theory.  The increase of $\rho$ with $x$ in cuprate superconductors is  
therefore readily accounted for by theories\cite{lee} in which superconductivity 
is due to the condensation of Cooper pairs formed from holes in a doped Mott insulator\cite{anderson}.
Theories which start with this view must still explain the fact that  
$\rho$ vanishes at a non-zero value of $x$, and deal with the awkward property 
that cuprate superconductivity evolves smoothly from the underdoped regime 
to an overdoped regime in which it appears to be
explainable in terms of conventional band-quasiparticle Cooper pair condensation.
In this article we propose an alternate explanation for the $x$-dependence of $\rho$ 
based on band-quasiparticle pairing.  Our argument accounts for the correlation energy of a d-wave 
superconductor in the presence of incipient antiferromagnetism and is based on 
the following general expression for the phase stiffness of a superconductor:
\begin{equation}
\rho = \frac{1}{A} \frac{d^{2} E}{d P^2},
\end{equation}
where $A$ is the area of the system, $\vec{P}$ the pairing wavevector\cite{phasegradient}, and 
$E$ is the total energy including both mean-field and correlation contributions: $E=E^{MF}+E^{cor}$.
The familiar BCS theory expression for $\rho$ captures only the mean-field theory contribution to
the energy.

When superconductivity is viewed as a weak-coupling instability of a Fermi liquid, it is 
usually implicitly assumed that $E^{cor}$ is not significantly influenced by the formation 
of the superconducting condensate, and certainly not by changes in the condensate's pairing momentum $\vec{P}$.
In the case of simple models with parabolic bands and galilean invariance, neglect of the 
correlation energy contribution can be justified rigorously.
We argue the correlation energy contribution is significant in underdoped cuprates because there is
direct competition between the Fermi sea quantum fluctuations which condense in 
antiferromagnetic and d-wave superconducting states.  Consequently the pair-breaking 
effects of finite $\vec{P}$, which weaken superconductivity, also increase the 
importance of antiferromagnetic fluctuations, lowering $E^{cor}$ and 
decreasing $\rho$ compared to its mean-field value.  In the following sections we first use  
a fully phenomenological and then a partially microscopic extended-Hubbard-model weak-coupling theory to expand on this
idea.  The conjugate relationship\cite{demler} between pairing and antiferromagnetic fluctuations 
plays an important role in the fluctuation spectrum and hence in theories of the correlation energy.
In our theory of the underdoped state, the resonant magnetic mode (INSR) observed in inelastic neutron
scattering\cite{mook,keimer} experiments therefore has a somewhat different interpretation than
in most earlier theory\cite{resonancemodeearly,resonancemoderpa,resonancemodepre,tch}, appearing as a 
kind of magnetic plasmon.    

\section{Phenomenological Theory}

The basic ideas of our theory are qualitative,
independent of most microscopic details, and most easily described  
in terms of the properties of a low-energy effective-field 
model for the collective fluctuations of a weak-coupling d-wave 
superconductor.  The relationship to less transparent
generalized random-phase approximation (GRPA) correlation energy calculations is explained below.
We construct a quantum action by introducing a 
set of states which incorporate the coupled triplet-pairing and spin-density fluctuations on which we focus.
$\vert \Psi[\phi,V]\rangle$ is the Fock-space Slater determinant  
ground state of the quadratic Hamiltonian
\begin{equation}
\begin{array}{ll}
\displaystyle
{\cal H}^{fluc}=&{\cal H}_{MF} + \sum_{i\sigma} \sigma V_i \; c_{i\sigma}^{\dagger} c_{i\sigma}\\[2mm]
\displaystyle
&+ \Delta_0 \big[ \sum_{i\tau} (-)^{\tau} [\exp(i\phi_i)-1] c_{i\uparrow}^{\dagger} c_{i+\tau\downarrow}^{\dagger} 
 + h.c. \big]. 
\end{array}
\label{h1body}
\end{equation} 
(For notational simplicity we have exhibited here only 
fluctuations with zero spin projection along the quantization direction.) 
In Eq.[~\ref{h1body}], $\tau$ labels the four neighbours of each site on a two-dimensional
square lattice, and $(-)^{\tau}$ represents the d-wave variation of mean-field near-neighbor pair
potentials.
%$\phi_i$ is independent of $\tau$ when near-neighbo.  
Using these states as an approximate identity resolution
leads to the following low-energy imaginary-time action 
for the collective variables $\phi_{i}$ and $V_{i}$:
\begin{equation}
{\cal S} = \int_{0}^{\infty} \; d\tau \Big[\; \hbar \; \langle \Psi[\phi,V] \vert \partial_{\tau} \vert \Psi[\phi,V]
\rangle + E[\phi,V]\; \Big],
\label{colaction}
\end{equation}
where $E[\phi,V]= \langle \Psi[\phi,V] \vert {\cal H} \vert \Psi[\phi,V] \rangle$ and ${\cal H}$ is the 
full microscopic Hamiltonian.  Mean-field theory states are obtained by minimizing $E[\phi,V]$.
The first term in the action captures the Berry phase coupling\cite{demler} between 
pairing and spin-density fluctuations on which we now elaborate.

\begin{figure}
\includegraphics{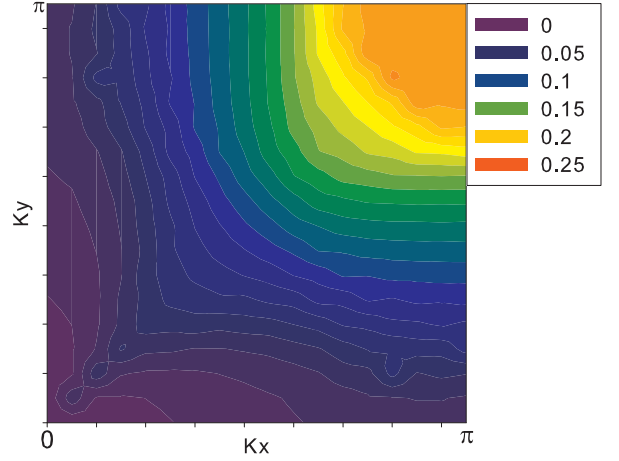}
\caption{\label{fig:one} (Color online) Berry's curvature $C_{\vec{k}}$ {\em vs.} $\vec{k}$ for the $d$-wave mean-field state of  
a generalized Hubbard model with $U/t=2.0, V/t=2.0, t'/t=-0.3$, and $x=0.12$. $
\Delta = \langle c_{i\uparrow}c_{j\downarrow}\rangle = (-)^\tau 0.145$ in this case.}
\end{figure}

The potentials associated with the two types of fluctuations are:
\begin{eqnarray} 
\partial {\cal H}^{(fluc)}/\partial V_{\vec k} &=& \sum_{\sigma,\vec{p}} \sigma
c_{\vec{p}-\vec{k},\sigma}^{\dagger} c_{\vec{p},\sigma} \nonumber \\
\partial {\cal H}^{(fluc)}/\partial \phi_{\vec{k}} &=& i \sum_{\vec{p}} \Delta_{\vec{p}}
\big[ c_{\vec{p}-\vec{k},\uparrow}^{\dagger} c_{\vec{-p},\downarrow}^{\dagger}  - h.c. \big].
\label{forces} 
\end{eqnarray} 
The Berry phase term can be evaluated explicitly for small fluctuations
by using perturbation theory expressions for the wavefunctions 
which appear in the Slater determinant $\vert \Psi[\phi,V]\rangle$:
\begin{equation}
{\cal S}_{Berry} = \int_{0}^{\infty} d \tau \sum_{\vec {k}} \; C_{\vec{k}} \;  \phi_{-\vec{k}}\partial_{\tau} V_{\vec k},
\end{equation}
where
\begin{equation}
C_{\vec k} = 2 \sum_{\vec{p}} \frac{ {\rm Im}\Big[
\langle \chi_{\vec p,-}|\frac{\partial {\cal H}^{fluc}}{\partial V_{\vec k}} | \chi_{\vec{p}+\vec{k},+}\rangle
\langle \chi_{\vec{p}+\vec{k},+}|\frac{\partial {\cal H}^{fluc}}{\partial \phi_{-\vec{k}}} |\chi_{\vec{p},-}\rangle \Big]} 
{(E_{\vec{p}+\vec{k}}+E_{\vec{p}})^{2}}.
\label{curvature}
\end{equation}
In Eq.(~\ref{curvature}) we have made the usual Nambu spin-down particle-hole transformation of the mean-field Hamiltonian
so that it has two eigenstates at each wavevector in the square lattice Brillouin zone with eigenvalues
$\pm E_{\vec{p}}$, one ($\chi_{\vec{p},-}$) occupied and one ($\chi_{\vec{p},+}$) unoccupied. 
In Fig.[~\ref{fig:one}] we show Berry curvature values calculated 
from this expression as a function of $\vec{k}$ which are strongly peaked
near $\vec{k} = \vec{Q}=(\pi/a,\pi/a)$; these results are robust over a broad range of dopings, gap sizes, and 
band-structure models.  Pairing phase fluctuations are conjugate to spin-density fluctuations for $\vec{k}$ near 
$\vec{Q}$, just as they are conjugate to charge-density fluctuations for $\vec{k}$ near $0$,
because of\cite{demler} the d-wave property $\Delta_{\vec{p}+\vec{Q}}= - \Delta_{\vec{p}}$. 

We now argue that there is competition between the correlation energy gain 
due to antiferromagnetic fluctuations and d-wave singlet Cooper pair formation.
Strong experimental evidence for this competition
is provided by the apparent enhancement\cite{lee,other} of antiferromagnetic fluctuations 
in cuprate vortex cores and in cuprates placed in an external magnetic fields.
Changes in external conditions which weaken superconductivity enhance antiferromagnetism. 
Here we explore consequences of this competition for the 
correlation contribution to the superfluid density, 
{\em i.e.} for the dependence of correlation energy on pairing-momentum 
$\vec{P}$.

In our model the quadratic fluctuation action of a d-wave superconductor is   
\begin{equation}
\begin{array}{l}
\displaystyle
{\cal L}_{fluc} =  \frac{1}{2 \beta} \sum_{\omega,\vec{k}}
\big[- 2 \hbar \omega \; C_{\vec{k},\omega} \; V(-\vec{k},-\omega) \; \phi(\vec{k},\omega) \\[2mm]
\displaystyle
 \; \; + \; K^{sp}_{\vec{k},\omega} \;
 |V(\vec{k},\omega)|^2 \; + \; K^{\phi}_{\vec{k},\omega} \; |\phi(\vec{k},\omega)|^2 \; 
\big]. 
\end{array}
\label{fluclagrangian} 
\end{equation}
In Eq.[~\ref{fluclagrangian}],
$K^{\phi}$ and $K^{sp}$ are phase and spin-density
stiffnesses. The onset of antiferromagnetism
occurs when $K^{sp}_{\vec{Q},\omega=0}=0$.  
In using this action we assume that the most important quantum fluctuations are 
d-wave pair phase and spin-fluctuations.  The microscopic calculations GRPA described
in the next section support this assumption. 
In Eq.[~\ref{fluclagrangian}], frequency dependence is indicated in $C_{\vec{k}}$, $K^{sp}_{\vec{k}}$, and $K^{\phi}_{\vec{k}}$ 
to recognize the existence of non-adiabatic effects accounted for in these more microscopic calculations but neglected in 
this qualitative discussion.
%(These become increasingly important as $x$ approaches the optimally and overdoped regimes.)  
The quadratic fluctuation action  
then describes a system with collective modes at energies
\begin{equation}
E^{res}_{\vec{k}} = \frac{\sqrt{K^{sp}_{\vec{k}} \; K^{\phi}_{\vec{k}}}}{C_{\vec{k}}},
\end{equation}
and a corresponding zero-point energy contribution 
\begin{equation}
E^{zp} = \sum_{k}' E_{\vec{k}}/2.
\end{equation}
This adiabatic theory of the INSR mode is accurate only 
when $E^{res}_{\vec{k}}$ lies below the particle-hole continuum; the prime on the wavevector 
sum above recognizes that this condition is satisfied only near $\vec{k}=\vec{Q}$. 
The fluctuation correction to $\rho$ can be related to the pairing-wavevector dependence 
of the zero-point energy as we explain below. 
 
We expect $K^{sp}(\vec{Q})$ to decrease with $\vec{P}$ because suppressed 
pairing favors antiferromagnetism.  The strength of this dependence can be 
estimated roughly from experiment by associating the magnetic length $\ell_B$ at the 
magnetic field strength required to induce antiferromagnetism in a cuprate superconductor with the value of $P^{-1}$ 
at which $K^{sp}(\vec{Q})$ goes to zero.  Taking a typical value for 
this field $\sim 100 \, {\rm Tesla}$ and assuming that the 
resonance mode is well defined over the portion of the BZ 
with large Berry curvature (say $\sim 10\%$), gives a negative correlation energy contribution to
the phase stiffness per two-dimensional cuprate layer of  
$\rho^{cor} \sim - 0.1 n \ell_B^2 \, E^{res} \sim - E^{res} \sim 0.05 {\rm eV}$, comparable to the 
value of $\rho$ inferred from penetration depth measurements in optimally doped samples.
Although this estimate is clearly 
very rough, it does establish that the correlation correction to $\rho$ can be substantial in the
underdoped regime. 

Charge density fluctuations are not included in this analysis because their Berry phase coupling
to the phase fluctuations of the superconducting order parameter is large only
near $\vec{k}=0$ and negligible near $\vec{k}=\vec{Q}$\cite{nagaosa}. 
Thereofore the charge density fluctuations do not play a significant roles for the physics near 
$\vec{k}=\vec{Q}$, instead they cause the instabilities near $\vec{k}=0$ in the microscopic GRPA calculations which we mention
below.

\section{Microscopic GRPA Theory}
We now evaluate the correlation energy of an extended Hubbard model\cite{tuvrev} in the generalized random 
phase approximation (GRPA) approximation.  The model we study in this article has on-site repulsive interactions $U$ which 
drive antiferromagnetism and near-neighbour attractive interactions $V$ which drive d-wave superconductivity: $H=H_t+H_U+H_V$,
\begin{equation}
\begin{array}{l}
\displaystyle
H_t=-t\sum_{<i,j>,\sigma} c^\dagger_{i\sigma} c_{j\sigma} + h.c. -t'\sum_{<i,j>',\sigma} c^\dagger_{i\sigma} c_{j\sigma} + h.c.\\[2mm]
\displaystyle
H_U=U\sum_i \hat{n}_{i\uparrow}\hat{n}_{i\downarrow}\,\,\,,\,\,\,H_V=-V\sum_{<i,j>\sigma\sigma'}\hat{n}_{i\sigma}\hat{n}_{j\sigma'}.
\end{array}
\label{tuv}
\end{equation}
In Eq.(~\ref{tuv}) $U$, $V$, $t$ and $t'$ should all be thought of as effective parameters which apply at the 
energy scale of pairing and depend on $x$.  Values for $V$, $t$, and $t'$ can be estimated from ARPES data\cite{arpes}.
Spin dependent Heisenberg near-neighbor interactions of the type used in $t-J$ models could also be used in 
the low-energy effective Hamiltonian, but are neglected here for simplicity.  The conclusions we draw in this article
do not depend on whether the near-neighbor effective interaction which drives d-wave superconductivity is 
spin-independent or spin-dependent.

The GRPA correlation energy of a $d$-wave condensate state with pairing momentum $\vec{P}$
is\cite{tdhft}:
\begin{equation}
E^{cor}(\vec{P})=\frac{1}{2}\, \sum_{\vec{q},i}\;  \big[\, \omega_{i}(\vec{P},\vec{q})- \epsilon^{ph}_i(\vec{P},\vec{q}) \big],
\end{equation}
where $\epsilon^{ph}_i(\vec{P},\vec{q})$ is a quasiparticle particle-hole exictation 
energy, and $\omega_i(\vec{P},\vec{q})$ is the corresponding GRPA excitation energy.
This equation can be derived by expanding the 
GRPA Hamiltonian to quadratic order in particle-hole excitation amplitudes approximated  
as independent bosons.  The correlation energy expression then drops out of a 
boson Bogoliubov transformation.  The analysis in the preceeding qualitative discussion
assummed that the $\vec{P}$-dependence of $E^{cor}$ is dominated by its collective mode contribution,
an assumption that is largely justified by the following more microscopic calculation.
 
In a GRPA theory excitation energies $\omega_{i}(\vec{P},\vec{q})$ are obtained from a time-dependent mean-field-theory\cite{tdhft}
in which the quasiparticles respond to the external potential and to induced mean-field potentials:
$H'=H^{ext}+H^{f}(t)$ where $H^{f}(t)=H^{f1}(t)+H^{f2}(t)$,
\begin{equation}
\begin{array}{ll}
\displaystyle
H^{f1}(t)=\frac{1}{A}\sum_{\vec{p},\vec{k},\vec{q},\sigma} &F(\vec{q})\left[\delta\langle c^\dagger_{\vec{p}+\vec{q}\bar{\sigma}}\,c_{\vec{p}\bar{\sigma}}\rangle 
c^\dagger_{\vec{k}-\vec{q}\sigma}\,c_{\vec{k}\sigma}\right]\\[2mm]
\displaystyle
&+G(\vec{k},\vec{p},\vec{q})\left[\delta\langle c^\dagger_{\vec{p}+\vec{q}\sigma}\,c_{\vec{p}\sigma}\rangle c^\dagger_{\vec{k}-\vec{q}\sigma}\,c_{\vec{k}\sigma}\right]\\[2mm]
\displaystyle
&+H(\vec{k},\vec{p})\left[\delta\langle c^\dagger_{\vec{q}-\vec{p}\bar{\sigma}}\,c^\dagger_{\vec{p}\sigma}\rangle c_{\vec{k}\sigma}\,c_{\vec{q}-\vec{k}\bar{\sigma}}+h.c.\right],
\\[2mm]
\displaystyle
H^{f2}(t)=\frac{1}{A}\sum_{\vec{p},\vec{k},\vec{q},\sigma}&I(\vec{k},\vec{p})\left[\delta\langle c^\dagger_{\vec{p}\sigma}c_{\vec{p}-\vec{q}\bar{\sigma}}\rangle
c^\dagger_{\vec{k}-\vec{q}\bar{\sigma}}c_{\vec{k}\sigma}\right]\\[2mm]
\displaystyle
&+J(\vec{k},\vec{p})\left[\delta\langle c^\dagger_{\vec{q}-\vec{p}\sigma}c^\dagger_{\vec{p}\sigma}\rangle
c_{\vec{k}\sigma}c_{\vec{q}-\vec{k}\sigma} + h.c.\right],\\[2mm]
\end{array}
\label{hflu}
\end{equation}
where
\begin{equation}
\begin{array}{l}
\displaystyle
F(\vec{q})=U-2V[\cos q_x+\cos q_y],\\[2mm]
\displaystyle
G(\vec{k},\vec{p},\vec{q})=2V[\cos(k_x-p_x-q_x)+\cos(k_y-p_y-q_y)\\[2mm]
\,\,\,\,\,\,\,\,\,\,\,\,\,\,\,\,\,\,\,\,\,\,\,\,\,\,\,\,\,\,-\cos q_x-\cos q_y],\\[2mm]
\displaystyle 
H(\vec{k},\vec{p})=U/2-V[\cos(k_x-p_x)+\cos(k_y-p_y)],\\[2mm] 
\displaystyle
I(\vec{k},\vec{p})=[-U+2V(\cos(k_x-p_x)+\cos(k_y-p_y))],\\[2mm]
\displaystyle 
J(\vec{k},\vec{p})=-V[\cos(k_x-p_x)+\cos(k_y-p_y)],
\end{array}
\end{equation}
and $\bar{\sigma}=-\sigma$. 
Because the d-wave BCS ground state is a spin-singlet, its elementary excitations consist of $S=1$ triplet
and $S=0$ singlet branches. 
$H^{f2}(t)$ captures the $S_z = \pm 1$ portions of the triplet fluctuations, studied by Demler and Zhang in a different context\cite{demler}. 
$H^{f1}(t)$ captures singlet and triplet $S_z=0$ fluctuations. 

Applying linear response theory to the quasiparticle Hamiltonian 
$H=H_{MF}(\vec{P}) + H'(t)$, we can compute the change in element of density matrix $\delta\langle\rho_{ab}\rangle$ by:
\begin{equation}
\delta\langle\rho_{ab}(t)\rangle=\frac{i}{\hbar}\int_{-\infty}^\infty dt'\theta(t-t')\,\langle [H'(t'), \rho_{ab}(t)] \rangle_{H_{MF}},
\label{lrt}
\end{equation}
, leading to an equation of the form:
\begin{equation}
(\omega\hat{I}-\hat{M})\bar{\rho}(\vec{q},\omega)=H^{ext},
\end{equation}
where $\bar{\rho}$ is a column representing the change in the quasiparticle density matrix.
The collective mode energies $\{\omega_i(\vec{q})\}$ are the eigenvalues\cite{caveatposneg} of the matrix
$\hat{M}$ which can be read off Eq.(~\ref{lrt}).
% PRB Comment.  Do we want to list explicit expressions for the matrix M here. I 
% beleive that we should since we have the room.
 For a given pairing momentum 
$\vec{P}$ and excitation momentum $\vec{q}$, the number of particle hole pairs is proportional to the number 
of momenta in the Brillouin-zone, which we limit by using periodic boundary conditions with a finite quantization area $A=L^2$.
Diagonalizing $\hat{M}$ is equivalent to performing the boson Bogoliubov transformation, and 
equivalent to summing the ladder and bubble diagrams used to represent the GRPA in diagramatic perturbation theory.

\begin{figure}
\includegraphics{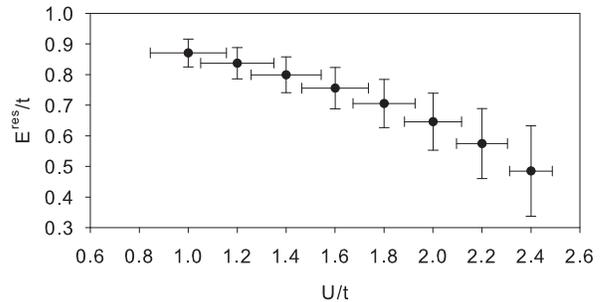}
\caption{\label{fig:two} Resonance mode energy and weights in spin and pair response
functions {\em vs.} $U$ for $\vec{q}=\vec{Q}$ and doping $x=0.12$.  These results were obtained for
$\vec{P}=0$ and calculated with $V/t=2.0$ and $t'/t=-0.3$ and $34 \times 34$ $\vec{k}$ points in the BZ.  For each value of $U$
the vertical (horizontal) bar represents the collective mode weight in the spin
(pair) response function. The weights were evaluated using the same response function defintions
as Tchernyshyov {\it et. al.}\cite{tch}.}
\end{figure}

% PRB Comment:  I wonder if we should add a new schematic figure and the additional
% results we have discussed which explain very clearly why finite P lowers the correlation energy.
% This is partially easier to do with the material we have developed in the other paper - but
% what can we do here?  Perhaps if we add a plot of the dependence of \Delta on 
% momentum - maybe that's the main effect i.e. \chi^{S} grows because \Delta decreases ? 
%  
Typical GRPA extended Hubbard model results for the spin and pair response functions
of the $\vec{P}=0$ state at wavevector $\vec{q}=\vec{Q}$ exhibit a single collective mode below the particle-hole continuum 
with large weight in both responses.  In Fig.[~\ref{fig:two}],we illustrate the dependence 
on $U$ of the energy of this excitation, and of its weight 
in spin-density and triplet-pair response functions, with 
$V$, $t$ and $t'$ (and hence the d-wave mean-field state) held fixed.  
As $U$ increases, promoting antiferromagnetism, the collective mode excitation energy 
decreases, its weight in the spin response function increases, and its weight in the 
pair-excitation spectrum decreases.  These properties are consistent with our qualitative 
effective theory, given the expectation that $K^{sp}_{\vec{Q}}$ should decrease with $U$. 
The opposing variations of triplet-pair and spin-density weights demonstrates that the Berry 
phase mechanism dominates coupling between antiferromagnetic and pairing phase fluctuations as expected.

\section{ Discussion} 

\begin{figure}
\includegraphics{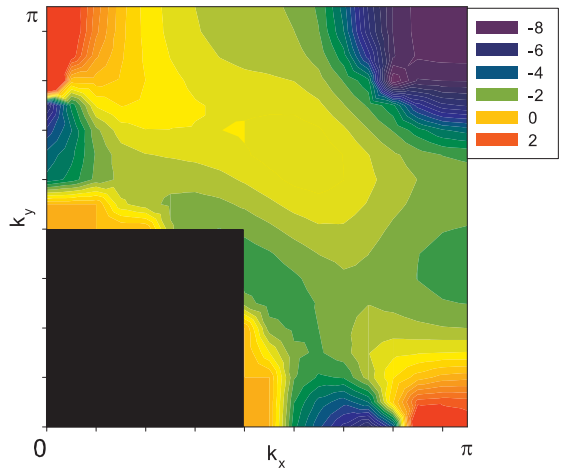}
\caption{\label{fig:three} (Color online) Correlation contribution to $\rho$ from fluctuations with wavevector $\vec{k}$
($\,\rho^{cor}(\vec{k})$ for the same model parameters as in Fig.[~\ref{fig:one}].  $\rho^{cor}(\vec{k})$ is normalized so that its Brillouin-zone average
is $\rho^{cor}/\rho^{MF}$, which has the value $-1.6$ for these parameters.
The GRPA excitation energies in the black area near $\vec{k}=(0,0)$ are imaginary reflecting longer length scale
instabilities\cite{zaanen} of the extended Hubbard model we use.  These long-wavelength instabilities are sensitive
to model details and independent of the $\rho^{cor}$ contributions from near $\vec{k}={\pi,\pi}$.}
\end{figure}

Our qualitative discussion suggested that there should be a strong negative correlation
contribution to $\rho$ because the INSR near $\vec{q}=\vec{Q}$ softens with increasing $|\vec{P}|$. 
Because we perform our calculations with periodic boundary conditions, 
we estimate the mean-field ($\rho^{MF}$) and correlation ($\rho^{cor}$) contributions 
to $\rho$ by comparing energies at $\vec{P}=0$ and $\vec{P}=\vec{P}^{min}=(2\pi/L,0)$;  
$\rho \approx [E(\vec{P}^{min}) - E(0)]/2\pi^2$.
Fig.[~\ref{fig:three}] illustrates the fluctuation-wavevector dependence of  
correlation contributions to $\rho$.  As expected we find that\cite{longpaper}
modes near $\vec{q}=\vec{Q}$ soften, making a negative contribution to $\rho$. 
This dependence of collective mode energies on the pair-momentum of the superconducting condensate is unusual
and is indicative of the microscopic competition between antiferromagnetism and d-wave superconductivity. 
This result contrasts strongly with the absence of any significant dependence of plasmonic collective modes on 
pair condensate properties in conventional superconductors. 

As indicated in Fig.[~\ref{fig:three}] we also find that for the model parameters chosen,
collective modes at momenta near $(0,0)$ have complex energies for both values of $\vec{P}$.  
This finding reflects the tendency of extended Hubbard models, and of real cuprate 
materials, to longer period density-wave instabilities\cite{zaanen}.
We do not believe that these ubiquitous instabilities, which appear to be material specific, 
should not play an essential role in underdoped-cuprate superfluid density suppression since
long-wavelength density-wave order will have little impact on near-neighbour 
antiferromagnetic fluctuations.   

We find that $\rho^{cor}/\rho^{MF}$ is negative and of order $-1$ in the underdoped regime 
when extended Hubbard model parameters are in the range thought to represent underdoped cuprates. 
We conclude that a substantial suppression of the superfluid density  
due to the pairing wavevector dependence of the correlation energy occurs in 
underdoped cuprates and that it is responsible for the downturn in the critical temperature. 
Our weak-coupling theory is unable to describe physics very near the termination of
superconductivity on the underdoped side, although there is some indirect evidence from experiment\cite{ubc,ohiostate}
(for example from the relatively weak temperature dependence of $\rho$) that critical 
fluctuations are important in a relatively narrow doping range.  
Our explanation for reduced superfluid density in 
underdoped cuprates is independent of the microscopic origin of the effective 
near-neighbor interaction responsible for $V$ and hence for d-wave superconductivity.

\section{Acknowledgments}
This work was supported by the Welch Foundation, the National Science Foundation under
grants DMR-0606489 and DMR-0547875, and a University of Waterloo start-up grant.  
The authors acknowledge helpful interactions with Dan Arovas, Andrey Chubukov, Eugene Demler, 
Bernhard Keimer, Patrick A. Lee, M.R. Norman, and Oleg Tchernyshyov.

\end{document}